\documentclass[12pt]{elsart}
\usepackage{amsfonts,amssymb,amstext}
\catcode`\@=11
\def\@journal{{\tt hep-th/9511229}}
\catcode`\@=12
\def\ope[#1][#2]{\mathord{#2\over{\ifnum#1=1 {z-w} \else {(z-w)^{#1}}\fi}}}
\def\pair<#1,#2>{\mathop{\left\langle#1\mathbin{,}#2\right\rangle}\nolimits}
\def\fr#1#2{{\textstyle {#1\over#2}}}
\renewcommand{\d}{\partial}
\newcommand{\reg}{\text{reg}}
\renewcommand{\gg}{{\ensuremath{\mathfrak g}}}
\newcommand{\gh}{{\ensuremath{\mathfrak h}}}
\newcommand{\gk}{{\ensuremath{\mathfrak k}}}
\newcommand{\sG}{{\ensuremath{\mathsf G}}}
\newcommand{\sT}{{\ensuremath{\mathsf T}}}
\newcommand{\sJ}{{\ensuremath{\mathsf J}}}

\newcommand{\W}{{\ensuremath{\mathsf W}}}
\newcommand{\NPB}[3]{{\sl Nucl. Phys.} {\bf B#1} (#2) #3}
\newcommand{\CMP}[3]{{\sl Comm. Math. Phys.} {\bf #1} (#2) #3}
\newcommand{\PRD}[3]{{\sl Phys. Rev.} {\bf D#1} (#2) #3}
\newcommand{\PLB}[3]{{\sl Phys. Lett.} {\bf #1B} (#2) #3}

\newcommand{\PTP}[3]{{\sl Prog. Theor. Phys.} {\bf B#1} (#2) #3}
\newcommand{\IJMPC}[3]{{\sl Int. J. Mod. Phys.} {\bf C#1} (#2) #3}
\begin{document}
\begin{frontmatter}
\title{$N{=}1$ and $N{=}2$ cosets from gauged supersymmetric WZW
models}
\author[QMW]{Jos\'e~M~Figueroa-O'Farrill\thanksref{EPSRC}\thanksref{emailjmf}}
and
\author[ICTP]{Sonia Stanciu\thanksref{emailss}}
\address[QMW]{Department of Physics, Queen Mary and Westfield College,
Mile End Road, London E1 4NS, UK}
\address[ICTP]{ICTP, P.O. Box 586, I-34100 Trieste, ITALY}
\thanks[EPSRC]{Supported in part by the EPSRC under contract
GR/K57824.}
\thanks[emailjmf]{\tt mailto:j.m.figueroa@qmw.ac.uk}
\thanks[emailss]{\tt mailto:sonia@ictp.trieste.it}
\begin{abstract}
We present a derivation of the $N{=}1$ and $N{=}2$ superconformal
coset constructions starting from a supersymmetric WZW model where a
diagonal subgroup has been gauged.  We work in the general framework
of self-dual (not necessarily reductive) Lie algebras; but even in the
reductive case these results are new.  We show that the BRST
cohomology of the gauged supersymmetric WZW model contains the $N{=}1$
(and if it exists also the $N{=}2$) coset generators.  We also extend
the BRST analysis to show that the BRST cohomology of the
supersymmetric WZW model agrees with that of an ordinary bosonic WZW
model (in a representation twisted by the presence of the coset
fermions).  In particular, in the case of the topological $G/G$ coset,
the supersymmetric and nonsupersymmetric theories agree.
\end{abstract}
\end{frontmatter}

\section{Introduction}

The relationship between the coset construction
\cite{earlycoset,GKO,FSN=0} $G/H$ and the gauged Wess--Zumino--Witten
(WZW) model is well-known.  The judicious use of the
Polyakov--Wiegmann identity and the careful consideration on the path
integral jacobians allows one to write the gauged WZW model in terms
of the original (ungauged) WZW model on $G$, an auxiliary WZW model on
$H$, and some anticommuting $H$-ghosts \cite{GWZW,FSN=0}.  Further
analysis of this gauge theory reveals that in the BRST cohomology one
has a realisation of the coset Virasoro algebra, and when $G$ is
compact, restricting to a suitable subclass of representations of the
affine Lie algebra, one recovers
\cite{GK} precisely the unitary series of Virasoro representations
expected from the construction in \cite{GKO}.

The supersymmetric situation is somewhat different.  The $N{=}1$ and
$N{=}2$ coset constructions were studied originally in
\cite{KazamaSuzuki,Schweigert}, where only reductive---that is,
semisimple $\times$ abelian---Lie algebras were considered.  However a
satisfactory derivation of these coset constructions from a
supersymmetric gauged WZW model was lacking.\footnote{Superconformal
field theories were obtained in \cite{Schnitzer} (see also
\cite{Nakatsu}) from gauged supersymmetric WZW models, but these
theories do not coincide with the $N{=}1$ coset constructions---they
have different central charges.}  Correct path integral constructions
were written down in \cite{Witten}, derived in \cite{Nojiri} (for a
particular model) and in \cite{Tseytlin} (in general).  However a
conformal field theoretical derivation of the supersymmetric coset
constructions from a gauged WZW model, in the style of \cite{GWZW},
does not exist.  The point of this letter is to remedy this situation.

This letter is organised as follows.  In Section 2 we briefly set the
notation and review the conformal field theory describing the $N{=}1$
supersymmetric WZW model and the gauging of a diagonal subgroup.  We
then describe the superconformal field theories described by these
models: the $N{=}1$ (affine) Sugawara construction in Section 3 and
the $N{=}1$ coset construction in Section 4.  In Section 5 we prove
that the natural $N{=}1$ Virasoro algebra of the gauged $N{=}1$ WZW
model is both BRST invariant and BRST cohomologous to the one coming
from the coset construction.  In Section 6 we discuss the $N{=}2$
cosets and we prove that when these cosets exist, the extra $N{=}2$
generators are also BRST invariant.  In Section 7 we prove that the
BRST cohomology of the gauged supersymmetric WZW model reduces to that
of a gauged bosonic WZW model coupled to the coset fermions.  In
particular, when discussing the topological gauging $G/G$, the two
theories---supersymmetric and bosonic---agree.

Finally a note on the notation.  Because this work arose as part of
the programme initiated in \cite{FSN=0} and continued in \cite{FSN=1}
on WZW models based on Lie groups which are not necessarily compact,
we work under the assumption that Lie algebras are self-dual, but not
necessarily reductive.  For nonreductive Lie algebras the notion of
{\em level\/} does not make sense and is replaced by that of an
invariant metric, which need not be proportional to any canonical
metric (as is the case in simple Lie algebras, for example).
Similarly, the dual Coxeter number is now replaced (roughly) by the
Killing form.

\section{The gauged supersymmetric WZW model}

The $N{=}1$ WZW model is defined classically by the action
\begin{eqnarray*}
(i\,\alpha)^{-1} I^S_\Omega[G] ={} &&
\int_{\Sigma_S}\pair<G^{-1}DG,G^{-1}{\bar D}G>\\
&& {} + \int_{B_S}\pair<G^{-1}\d_t G, \left[ G^{-1}DG,G^{-1}{\bar
D}G\right]>~,
\end{eqnarray*}
where $\Sigma_S$ is a super-Riemann surface, $B_S$ a supermanifold
with boundary $\Sigma_S$ and $G$ is a superfield whose
$\theta$-independent component takes values in a Lie group $\mathcal
G$ which we assume posses a bi-invariant metric
$\Omega$.\footnote{$\Omega$ is written as $\pair<-,->$ above to avoid
cluttering the notation.} Expanding into components and solving for
the auxiliary fields, the above model has a particularly simple
description in terms of a bosonic $\mathcal G$-valued field $g$ and
Majorana--Weyl fermions $\psi$ and $\bar\psi$ with values in the Lie
algebra $\gg$:
\begin{equation}
I^S_\Omega[G] = I_\Omega[g] + I_\Omega[\psi,\bar\psi;g]~,
\end{equation}
where $I_\Omega[g]$ is the bosonic WZW action defined in \cite{FSN=0},
and $I_\Omega[\psi,\bar\psi;g]$ is the action for Majorana--Weyl
fermions axially coupled to the bosonic currents.

The quantum theory is described by the path integral
\begin{equation}
Z = \int [dg][d\psi][d\bar\psi] e^{-I_\Omega[g] -
    I_\Omega[\psi,\bar\psi;g]}~,
\end{equation}
or equivalently, by an $N{=}1$ affine Lie algebra
\cite{dVKPR,abdallas,Fuchs,KacTodorov} with data $(\gg, \Omega)$,
where we let $\Omega$ also denote the invariant metric on $\gg$.  Such
Lie algebras are known as self-dual.  (Of course, the theory has both a
holomorphic and an antiholomorphic sector, but as usual we concentrate
on the holomorphic one.)  Fix once and for all a basis $\langle
X_a\rangle$ for $\gg$, relative to which $\Omega$ has components
$\Omega_{ab}$ and such that the structure constants are ${f_{ab}}^c$.
This $N{=}1$ affine Lie algebra is generated by currents $I_a(z)$ and
fermions $\psi_a(z)$, obeying the following OPEs:
\begin{eqnarray}
I_a(z) I_b(w) &=& \ope[2][\Omega_{ab}] + \ope[1][{f_{ab}}^c I_c(w)] +
\reg\nonumber\\
I_a(z) \psi_b(w) &=& \ope[1][{f_{ab}}^c \psi_c(w)] + \reg\nonumber\\
\psi_a(z) \psi_b(w) &=& \ope[1][\Omega_{ab}] +
\reg~.\label{eq:KacTodorov}
\end{eqnarray}
Because $\Omega$ is nondegenerate, we can decouple the fermions from
the affine currents.  Indeed, in terms of the modified currents:
\begin{equation}
J_a(z) \equiv I_a(z) - \half \Omega^{bd} {f_{ab}}^c
(\psi_c\psi_d)(z)\label{eq:modJ}
\end{equation}
the OPEs (\ref{eq:KacTodorov}) become
\begin{eqnarray}
J_a(z) J_b(w) &=& \ope[2][\Omega_{ab}-\half \kappa_{ab}] +
\ope[1][{f_{ab}}^c J_c(w)] + \reg\nonumber\\
J_a(z) \psi_b(w) &=& \reg\nonumber\\
\psi_a(z) \psi_b(w) &=& \ope[1][\Omega_{ab}] +
\reg~,\label{eq:Affine+Fermions}
\end{eqnarray}
where $\kappa_{ab} = {f_{ac}}^d {f_{bd}}^c$ is the Killing form on
$\gg$.  Since $\gg$ is not necessarily semisimple, $\kappa$ need not
be nondegenerate.  Nevertheless, as shown in \cite{FSN=0,FSSD},
$\Omega - \half\kappa$ will generically be nondegenerate.

Now let $\gh\subset\gg$ be a Lie subalgebra such that the restriction
$\Omega|_{\gh}$ of $\Omega$ to $\gh$ remains nondegenerate.  Assume
that we have chosen the basis for $\gg$ in such a way that a sub-basis
$\langle X_i\rangle$ is a basis for $\gh$.  The condition on $\gh$
means that $\Omega_{ij}$ is an invariant metric on $\gh$.  (In
particular, $\gh$ is also self-dual.)  In \cite{FSN=1} it is proven
that this is a necessary and sufficient condition for the existence of
the $N{=}1$ coset construction $\gg/\gh$.  In this letter we will see
that the diagonally gauged supersymmetric WZW model reproduces this
coset construction.

As shown in \cite{Nojiri,Tseytlin,FSN=1} gauging {\em both\/} the
fermionic and bosonic symmetries corresponding to the diagonal
subalgebra $\gh$ gives rise to the following path integral:
\begin{eqnarray*}
Z ={}&& \int [dg][d\psi][d\bar\psi][d\tilde
h][d\tilde\psi][d\bar{\tilde\psi}][db][dc][d\beta][d\gamma][d\bar
b][d\bar c][d\bar\beta][d\bar\gamma]\\
&&{} \times  e^{-I_\Omega[g] - I_\Omega[\psi,\bar\psi;g]}\,
e^{I_{\Omega}[\tilde h] +
I_{\Omega}[\tilde\psi,\bar{\tilde\psi};\tilde h]}\,
e^{-I_{\mathrm{gh}}[b,c,\bar b, \bar c, \beta, \gamma, \bar\beta,
\bar\gamma]}~,
\end{eqnarray*}
where $(b,c)$ and $(\bar b, \bar c)$ are the ghosts familiar from the
nonsupersymmetric gauged WZW model, and $(\beta,\gamma)$ and
$(\bar\beta, \bar\gamma)$ are the (bosonic) ghosts corresponding to
the gauged fermionic symmetry.  Notice that unlike in the
nonsupersymmetric case, the metric of the `auxiliary' $\gh$-sector
does not get shifted.  This is because the jacobian responsible for
the shift receives now an equal but opposite contribution from the new
fermionic sector.

We may also describe this quantum theory as a superconformal field
theory consisting of three sectors coupled by a constraint, where
again we omit the antiholomorphic sector:
\begin{itemize}
\item[$\bullet$] the original $N{=}1$ affine Lie algebra with data
$(\gg,\Omega)$;
\item[$\bullet$] a second $N{=}1$ affine Lie algebra with data
$(\gh,-\Omega|_\gh)$ generated by $\tilde I_i(z)$ and
$\tilde\psi_i(z)$ subject to the OPEs:
\begin{eqnarray}
\tilde I_i(z) \tilde I_j(w) &=& \ope[2][-\Omega_{ij}] +
\ope[1][{f_{ij}}^k \tilde I_k(w)] +
\reg\nonumber\\
\tilde I_i(z) \tilde\psi_j(w) &=& \ope[1][{f_{ij}}^k \tilde\psi_k(w)]
+ \reg\nonumber\\
\tilde\psi_i(z) \tilde\psi_j(w) &=& \ope[1][-\Omega_{ij}] +
\reg~;\label{eq:gaugedHsector}
\end{eqnarray}
\item[$\bullet$] a supersymmetric ghost system with fermionic
$(b_i,c^i)$ and bosonic $(\beta_i,\gamma^i)$ generators subject to the
OPEs:
\begin{displaymath}
b_i(z) c^j(w) = \ope[1][\delta_i^j] + \reg\qquad\hbox{and}\qquad
\beta_i(z) \gamma^j(w) = \ope[1][\delta_i^j] + \reg~.
\end{displaymath}
\end{itemize}

The ghost sector also carries a realisation of the $N{=}1$ affine Lie
algebra with data $(\gh,0)$.  Indeed, we can define
\begin{displaymath}
I_i^{\mathrm{gh}}(z) \equiv {f_{ij}}^k b_k c^j - {f_{ij}}^k \beta_k
\gamma^j\qquad\hbox{and}\qquad \psi_i^{\mathrm{gh}}(z) \equiv
{f_{ij}}^k \beta_k c^j~,
\end{displaymath}
which obey the OPEs
\begin{eqnarray}
I_i^{\mathrm{gh}}(z) I_j^{\mathrm{gh}}(w) &=& \ope[1][{f_{ij}}^k
I_k^{\mathrm{gh}}(w)] + \reg\nonumber\\
I_i^{\mathrm{gh}}(z) \psi_j^{\mathrm{gh}}(w) &=& \ope[1][{f_{ij}}^k
\psi_k^{\mathrm{gh}}(w)] + \reg\nonumber\\
\psi_i(z) \psi_j(w) &=& \reg~.\label{eq:ghostsector}
\end{eqnarray}
Notice that since the metric is zero, the ``fermions''
$\psi_k^{\mathrm{gh}}$ cannot be decoupled; yet this will not
represent any problem.

Tensoring the three sectors we obtain a realisation of the $N{=}1$
affine Lie algebra with data $(\gh,0)$ generated by the {\em total\/}
fields:
\begin{displaymath}
I_i^{\mathrm{tot}}(z) \equiv I_i(z) + \tilde I_i(z) +
I_i^{\mathrm{gh}}(z)\quad\hbox{and}\quad\psi_i^{\mathrm{tot}}(z)
\equiv \psi_i(z) + \tilde \psi_i(z) + \psi_i^{\mathrm{gh}}(z)~.
\end{displaymath}
The absence of central terms in the algebra generated by
$I_i^{\mathrm{tot}}(z)$ and $\psi_i^{\mathrm{tot}}(z)$ just reiterates
the fact that we have gauged a non-anomalous symmetry and implies that
the BRST charge\footnote{We use the $[-,-]_n$ notation for operator
product expansions.  The properties of these brackets are summarised
for example in \cite{GetzlerMT,Kris}.} $d=[j_{\hbox{\tiny BRST}},-]_1$,
defined by
\begin{equation}
j_{\hbox{\tiny BRST}} = (I_i+\tilde I_i) c^i - (\psi_i+\tilde\psi_i)
\gamma^i - {f_{ij}}^k \beta_k c^i \gamma^j - \half {f_{ij}}^k b_k c^i c^j
\end{equation}
squares to zero.  In fact, the first order pole $[j_{\hbox{\tiny
BRST}},j_{\hbox{\tiny BRST}}]_1$ actually vanishes.  It will be
convenient to introduce the linear combinations:~ $\psi_i^\pm \equiv
\psi_i \pm \tilde \psi_i$.

\section{The $N{=}1$ Virasoro algebras}

Associated with any $N{=}1$ affine Lie algebra based on a self-dual
Lie algebra, there is a supersymmetric Sugawara construction which
yields an $N{=}1$ Virasoro algebra.  For the algebra defined by
(\ref{eq:KacTodorov}) we define:
\begin{displaymath}
G_\gg \equiv \Omega^{ab} J_a \psi_b - \fr{1}{6} f^{abc}
\psi_a\psi_b\psi_c\qquad\hbox{and}\qquad
T_{\gg} \equiv \half\Omega^{ab} J_aJ_b + \half\Omega^{ab}
\d\psi_a\psi_b~,
\end{displaymath}
where $f^{abc} \equiv \Omega^{ad} \Omega^{be} {f_{de}}^c$.  Notice
that we have used the decoupled currents $J_a$ defined by equation
(\ref{eq:modJ}).  $G_{\gg}$ and $T_{\gg}$ satisfy an $N{=}1$ Virasoro
algebra with central charge $c_{\gg} = \fr{3}{2}\dim\gg - \half
\Omega^{ab}\kappa^{\gg}_{ab}$, where we now let $\kappa^{\gg}$ denote
the Killing form of $\gg$.

Similarly for the $N{=}1$ affine Lie algebra defined by
(\ref{eq:gaugedHsector}) one defines
\begin{displaymath}
\tilde G_\gh \equiv -\Omega^{ij} \tilde J_i \tilde\psi_j - \fr{1}{6} f^{ijk}
\tilde\psi_i\tilde\psi_j\tilde\psi_k\quad\hbox{and}\quad
\tilde T_{\gh} \equiv -\half\Omega^{ij} \tilde J_i \tilde J_j -
\half\Omega^{ij} \d\tilde\psi_i\tilde\psi_j~,
\end{displaymath}
where $\tilde J_i$ are the decoupled currents in this sector which are
defined similarly to those in equation (\ref{eq:modJ}) but with metric
$-\Omega_{ij}$.  The $N{=}1$ Virasoro algebra satisfied by $\tilde
G_\gh$ and $\tilde T_{\gh}$ has central charge $\tilde c_{\gh} =
\fr{3}{2}\dim\gh + \half \Omega^{ij}\kappa^{\gh}_{ij}$.

Finally we define the $N{=}1$ Virasoro algebra for the ghost system.
Since the fermionic ghosts have weights $(1,0)$ and the bosonic ghosts
have weights $(\half,\half)$, we write
\begin{displaymath}
G_{\mathrm{gh}} \equiv b_i\gamma^i + \beta_i\d c^i\qquad\hbox{and}\qquad
T_{\mathrm{gh}} \equiv -b_i \d c^i + \half \left( \beta_i\d\gamma^i -
\d\beta_i\gamma^i\right)~.
\end{displaymath}
The $N{=}1$ Virasoro algebra satisfied by $G_{\mathrm{gh}}$ and
$T_{\mathrm{gh}}$ has the expected central charge $c_{\mathrm{gh}} =
-3\dim\gh$.

Tensoring all three $N{=}1$ Virasoro algebras together we find a
realisation with total central charge
\begin{equation}
c_{\mathrm{tot}} \equiv c_{\gg} + \tilde c_{\gh} + c_{\mathrm{gh}} =
\fr{3}{2}\left(\dim\gg - \dim\gh\right) -
\half\left(\Omega^{ab}\kappa^{\gg}_{ab} -
\Omega^{ij}\kappa^{\gh}_{ij}\right)~,\label{eq:ctotal}
\end{equation}
which as we will see presently is the central charge of the $N{=}1$
coset construction.

\section{The $N{=}1$ coset construction}

Given a self-dual Lie algebra $(\gg,\Omega)$ and a Lie subalgebra
$\gh\subset\gg$ such that the restriction $\Omega|_{\gh}$ is
nondegenerate, we can define an $N{=}1$ coset construction.  This is
done as follows \cite{FSN=1}.  Define currents $\hat J_i \equiv I_i -
\half \Omega^{j\ell} {f_{ij}}^k \psi_k\psi_{\ell}$.  Notice that $\hat
J_i$ is not equal to the modified current defined in (\ref{eq:modJ})
(unless $\gg=\gh$) but that nonetheless the new currents are decoupled
from the $\gh$-fermions:
\begin{displaymath}
\hat J_i(z) \psi_j(w) = \reg~,
\end{displaymath}
and still define a realisation of an affine Lie algebra based on
$\gh$:
\begin{displaymath}
\hat J_i(z) \hat J_j(w) = \ope[2][\Omega_{ij} - \half
\kappa^{\gh}_{ij}] + \ope[1][{f_{ij}}^k \hat J_k(w)] + \reg~,
\end{displaymath}
where now $\kappa^{\gh}_{ij}$ is the Killing form on $\gh$, which need
not agree with the restriction to $\gh$ of the Killing form on $\gg$.

By assumption, the restriction of the metric $\Omega$ on $\gg$ to
$\gh$ is nondegenerate, so we can decompose $\gg = \gh \oplus
\gh^\perp$, which, because of the invariance of the metric, is not
just a decomposition of vector space but also one of $\gh$-modules.
If we let $\langle X_\alpha \rangle$ denote a basis for $\gh^\perp$,
we can summarise this discussion by saying that $\Omega_{i\alpha} = 0$
and that ${f_{i\alpha}}^j = 0$.  (Notice, however, that there is no
restriction on ${f_{\alpha\beta}}^\gamma$, hence the above
decomposition, although reductive, need not be symmetric.)

Define now the following $N{=}1$ Virasoro generators:
\begin{displaymath}
G_\gh \equiv \Omega^{ij} \hat J_i \psi_j - \fr{1}{6} f^{ijk}
\psi_i\psi_j\psi_k\qquad\hbox{and}\qquad
T_{\gh} \equiv \half\Omega^{ij} \hat J_i \hat J_j + \half\Omega^{ij}
\d\psi_i\psi_j~,
\end{displaymath}
whose central charge is given by $c_{\gh} = \fr{3}{2} \dim\gh - \half
\Omega^{ij} \kappa^{\gh}_{ij}$.

The $N{=}1$ coset theory is generated by
\begin{displaymath}
G_{\gg/\gh} \equiv G_{\gg} - G_{\gh}\qquad\hbox{and}\qquad
T_{\gg/\gh} \equiv T_{\gg} - T_{\gh}~.
\end{displaymath}
These fields obey an $N{=}1$ Virasoro algebra with central charge
$c_{\gg/\gh} \equiv c_\gg - c_\gh$ which agrees with
(\ref{eq:ctotal}), and, more importantly, they have regular OPEs with
$G_\gh$ and $T_\gh$.

\section{$N{=}1$ coset theory in BRST cohomology}

The equality between the central charge $c_{\gg/\gh}$ of the coset
construction and the central charge $c_{\mathrm{tot}}$ of the gauged
supersymmetric WZW model suggests that the two theories are actually
equivalent.  In fact, we now show that the BRST cohomology of the
gauged supersymmetric WZW model admits a realisation of the coset
theory.  Just like in the nonsupersymmetric case \cite{GWZW,FSN=0},
all we need to show is that the generators of the coset SCFT
$(G_{\gg/\gh}, T_{\gg/\gh})$ and of the gauged supersymmetric WZW SCFT
$(G_{\mathrm{tot}}, T_{\mathrm{tot}})$ are not just BRST-invariant but
also BRST-cohomologous, so that their differences
\begin{displaymath}
G' \equiv G_{\mathrm{tot}} - G_{\gg/\gh}\qquad\hbox{and}\qquad
T'\equiv T_{\mathrm{tot}} - T_{\gg/\gh}
\end{displaymath}
are BRST-exact.  Since the BRST operator $d$ is a derivation over the
operator product and since, under the operator product, it is
$G_{\mathrm{tot}}$ and $G_{\gg/\gh}$ which generate the algebra, it is
enough to show our claim on these generators.  Indeed, a short
calculation shows that both $G_{\mathrm{tot}}$ and $G_{\gg/\gh}$ are
BRST-invariant, and moreover that $G_{\mathrm{tot}} - G_{\gg/\gh} =
d\Theta$, with
\begin{displaymath}
\Theta = \half \Omega^{ij} b_i\psi^-_j + \half \Omega^{ij} \beta_i
(I_j - \tilde I_j) + \fr{1}{3} f^{ijk}\beta_i \left(\psi_j\psi_k +
\tilde\psi_j\tilde\psi_k - \psi_j\tilde\psi_k\right)~.
\end{displaymath}

\section{$N{=}2$ cosets}

Under certain circumstances the $N{=}1$ coset theory admits an extra
supersymmetry giving rise to an $N{=}2$ coset.  For $\gg$ a reductive
Lie algebra---that is, semisimple $\times$ abelian---this is the
celebrated Kazama--Suzuki construction
\cite{KazamaSuzuki,Schweigert,GetzlerMT}. For a general self-dual Lie
algebra $(\gg,\Omega)$ the conditions for the existence of an $N{=}2$
theory extending the $N{=}1$ coset $\gg/\gh$ are the following
\cite{FSN=1}.  Let $\gk\equiv \gh^\perp \subset \gg$ be the orthogonal
complement of $\gh\subset\gg$.  Then $\gk$ must possess {\em an
$\gh$-invariant, integrable complex structure compatible with the
restriction to $\gk$ of the metric $\Omega$}.  Comparing with
\cite{HullWitten}, this condition supports our intuition that the CFT
defined by the $N{=}2$ coset is indeed described (at least
classically) by a $\sigma$-model on the coset space $G/H$.

Let $A: \gk \to \gk$ denote the complex structure.  Relative to the
basis $\langle X_\alpha \rangle$ for $\gk$, $A$ has components
${A^\alpha}_\beta$.  Because $A$ is compatible with the metric
$A^{\alpha\beta} = - A^{\beta\alpha}$, where $A^{\alpha\beta} =
{A^\alpha}_\gamma \Omega^{\beta\gamma}$.  Define $\sJ$ and $\sG^2$ by
\begin{eqnarray*}
2i\,\sJ &\equiv& A^{\alpha\beta} \psi_\alpha\psi_\beta -
A^{\alpha\beta} {f_{\alpha\beta}}^c I_c\\
\sG^2 &\equiv& A^{\alpha\beta} J_\alpha\psi_\beta + \fr{1}{6}
A^{\alpha\alpha'} A^{\beta\beta'} A^{\gamma\gamma'}
f_{\alpha\beta\gamma} \psi_{\alpha'}\psi_{\beta'}\psi_{\gamma'}~.
\end{eqnarray*}
Then together with $\sG^1\equiv G_{\gg/\gh}$ and $\sT \equiv
T_{\gg/\gh}$, they obey an $N{=}2$ Virasoro algebra.

If the gauged supersymmetric WZW model is to describe the $N{=}1$
coset theory, any extended symmetry of the $N{=}1$ Virasoro algebra
which the coset theory admits, must be already present (maybe up
to BRST-exact terms) among the BRST-invariant fields in the WZW model.
Therefore we expect that the $N{=}2$ extension, whenever it exists,
must be BRST-invariant or, in this case, since they don't involve the
ghosts, actually gauge invariant.  Since $\sJ$ and $\sG^1$ generate
the rest of the $N{=}2$ Virasoro algebra, and $\sG^1$ is already
BRST-invariant, all we need to show is that $\sJ$ is BRST-invariant.
But this follows trivially from the $\gh$-invariance of the complex
structure.
This proves that the gauged supersymmetric WZW model does provide a
lagrangian realisation of the $N{=}2$ coset construction.

%
%

\section{Decoupling the $\gh$-fermions}

The structure of the BRST current $j_{\hbox{\tiny BRST}}$ is very
suggestive.  Notice that it can be written as the sum of two terms
whose charges separately square to zero.  Indeed, let us write
$j_{\hbox{\tiny BRST}} = j_0 + j_1$, where
\begin{displaymath}
j_0 = -\psi^+_i\gamma^i\quad\hbox{and}\quad
j_1 = (I_i+\tilde I_i)c^i - {f_{ij}}^k \beta_k c^i \gamma^j - \half
{f_{ij}}^k b_k c^i c^j~.
\end{displaymath}
The subscripts refer to the $(b,c)$ ghost number.  Given that both
$(b,c)$ and $(\beta,\gamma)$ ghosts numbers are separately conserved,
we have that the respective differentials $d_0 \equiv [j_0,-]_1$ and
$d_1 \equiv [j_1,-]_1$ form a double complex:
\begin{displaymath}
d_0^2 = d_1^2 = d_0 d_1 + d_1 d_0 = 0~.
\end{displaymath}
Moreover the form of $j_0$ is reminiscent of a Koszul complex.
Indeed, $(\psi^\pm_i, \beta_i,\gamma^i)$ forms a Kugo--Ojima quartet
(see, for instance, \cite{FKKO} for the relevant notions) and decouple
from the theory.  We can see this in either of two ways.

First of all, as in all double complexes, there is a spectral sequence
converging to the BRST cohomology whose $E_1$ and $E_2$ terms are the
cohomology $H_{d_0}$ of $d_0$ and $H_{d_1}(H_{d_0})$ respectively.
Since the cohomology of $d_0$ is isomorphic to the CFT obtained by
decoupling the above Kugo--Ojima quartet, the spectral sequence
degenerates at the $E_2$ term.  Thus the BRST cohomology is isomorphic
to the cohomology of $d_1$ acting on the remaining fields.

Alternatively, we can just change variables.  Let's introduce the
field
\begin{displaymath}
r(z) \equiv -\half \Omega^{j\ell} {f_{ij}}^k \beta_\ell \psi^-_k c^i~.
\end{displaymath}
Let $R = \oint r(z)$ denote its charge.  If $\phi(z)$ is any field, we
define the conjugation by $R$ as follows:
\begin{displaymath}
e^R\,\phi(z)\,e^{-R} \equiv \sum_{k\geq 0}  {1\over k!}
\underbrace{[r,\cdots [r,[r}_k,\phi(z)]_1]_1\cdots ]_1~.
\end{displaymath}
It is easy to see that with the $r(z)$ defined above, this sum is
actually finite.  Applying this conjugation to the BRST current, we
find
\begin{displaymath}
e^R\,j_{\hbox{\tiny BRST}}\,e^{-R} = j_0 +
j'~,\quad\hbox{where}\quad j' = (\hat J_i + \tilde J_i) c^i -
\half {f_{ij}}^k b_k c^ic^j~.
\end{displaymath}
Notice that now $j_0$ and $j'$ are completely decoupled, since $j'$
involves the currents $\hat J_i$ and $\tilde J_i$ which do not
interact with the $\gh$-fermions.  At the level of the fields,
conjugation by $R$ induces a change of variables which, as will be
shown in \cite{FSN=1}, factorises the path integral.  In the factor
involving the Kugo--Ojima quartet, the contribution from the bosonic
ghosts cancels precisely the contribution coming from the
$\gh$-fermions, {\em including the zero modes\/}: since all fields
$(\beta_i,\gamma^i)$ and $\psi^\pm_i$ have weight $\half$.  At the end
of the day, one is left with a theory of coset fermions $\psi_\alpha$
and a WZW model in which we have gauged the diagonal $\gh$ symmetry.
Notice that the coset fermions {\em are\/} gauged, since it is $\hat
J_i$ and not $J_i$ which appears in $j'$.  This theory is precisely
the starting point of \cite{Witten}.

As a consequence we have two equivalent descriptions of the physical
sector of the supersymmetric gauged WZW model (or equivalently of the
$N{=}1$ coset theory):
\begin{itemize}
\item[$\bullet$] the cohomology of $d$ acting on the superconformal
field theory generated by $(I_a,\tilde I_i, \psi_a, \tilde\psi_i,
b_i,c^i,\beta_i,\gamma^i)$; and
\item[$\bullet$] the cohomology of $d' \equiv [j',-]_1$ acting on the
superconformal field theory generated by $(\hat J_a, \tilde J_i,
\psi_\alpha, b_i,c^i)$.
\end{itemize}
When $\gh=\gg$, there are no coset fermions, and the gauged
supersymmetric WZW model reduces to an ordinary bosonic gauged WZW
model.

Finally let us remark that from the results in \cite{FMech}---where
the above operator $R$ was introduced---it follows that any given
bosonic gauged WZW model embeds into a supersymmetric gauged WZW model
in such a way that their BRST cohomologies are the same.  What we have
proven above is that in the case of the topological gauging $\gh=\gg$,
the converse also holds. In other words, the bosonic and
supersymmetric $G/G$ gauged WZW models are equivalent.

\begin{ack}
Some of this work was presented by one of us (SS) at the collaboration
meeting {\em Superstrings and the Physics of Fundamental
Interactions\/} held in London on October 30-31, 1995.  SS would like
to thank the organisers of the meeting for the invitation to speak and
the members of the String Theory group of Queen Mary and Westfield
College and Chris Hull in particular for the hospitality.  We
benefited once more from the excellent {\em Mathematica\/} package
{\em OPEdefs}, written by Kris Thielemans \cite{OPEdefs}.  We are
grateful to Jaume Roca for a careful reading of the manuscript.

After most of this work was finished we learnt from Henric Rhedin
\cite{Rhedin} that he has independently obtained some of the results
pertaining to the (reductive) $N{=}1$ coset theory.  It is a pleasure
to thank him for conversations on this and other related topics.
\end{ack}

\end{document}